\documentstyle [12pt,epsf] {article}
\newcommand{\gf}{$\gamma_5$}

\begin{document}
\bibliographystyle{unsrt}

\pagestyle{empty}

\rightline{\vbox{
	\halign{&#\hfil\cr
	&BNL-61617\cr
	&March 1995\cr}}}
\vskip 1in
\begin{center}
{\Large\bf
{Spurious Anomalies in Dimensional Renormalization}}
\vskip .5in
\normalsize
T.L.\ Trueman \footnote{This manuscript has been authored
under contract number DE-AC02-76CH00016 with the U.S. Department
of Energy.  Accordingly, the
U.S. Government retains a non-exclusive, royalty-free license to
publish or reproduce the published form of this contribution, or
allow others to do so, for U.S. Government purposes.}\\
{\sl Physics Department, Brookhaven National
Laboratory, Upton, NY 11973}
\end{center}
\vskip 1.5in
\begin{abstract}
A set of procedures is given for avoiding the spurious anomalies that are
generated when the 't Hooft - Veltman definition of \gf\ is used in conjunction
with renormalization by minimal subtraction. These procedures are derived
from the
standard procedure, which requires in addition various finite
renormalizations to
remove spurious violations of chiral symmetry. They apply to open fermion
lines,
including flavor changing currents, to closed fermion loops, including those
which contain true anomalous currents, and to anomalous loops connected to open
fermion lines, to all orders in QCD. \end{abstract}
\vfill \eject \pagestyle{plain}
\setcounter{page}{1}
\section{Introduction}

	Since 1972 when 't Hooft and Veltman \cite{HV} and others \cite{others}
introduced the  dimensional regularization of Feynman diagrams the
technique has
come to play a central role in perturbative field theory calculations.
Coupled with the dimensional renormalization procedure \cite{GH} it
has led to a relatively simple and systematic way of removing the
divergences that inevitably accompany such calculations and of
obtaining precise predictions for experimental quantities in terms of
well-defined theoretical quantities like coupling constants.
	In their original paper\space 't Hooft and Veltman recognized the
special character of $\gamma_{5}$ : it is an intrinsically four-
dimensional object, not zero-dimensional as habit often
leads us to assume. In that paper they gave a prescription for
handling $\gamma_5$ in any dimension; namely
\begin{equation}{
\gamma_5 = i\gamma_0\gamma_1\gamma_2\gamma_3\;} \; .
\label{eq:def}
\end{equation}
(This is often refered to as the HV prescription for short.)
Although this prescription has been very much used and studied and
generally shown to be consistent \cite{BM} \cite{JC}, it has
nevertheless proved vexing over the years and continues to do so.
	This definition treats the first four dimensions of space-time
asymmetrically from the rest so that
\begin{eqnarray}
 \{\gamma_5,\gamma_\mu\}=0 & \mbox{for $\mu=0,1,2,3$} \nonumber \\
    {[}\gamma_5,\gamma_\mu]=0 & \mbox{otherwise.} \label{eq:com}
\end{eqnarray}
This leads to a certain amount of algebraic complication when
compared with the ``naive'' $\{\gamma_5,\gamma_{\mu}\}=0$ for all $\mu$.
It turns out that in very many cases the same result is
obtained by either method, so people naturally prefer to be naive.
However, it was shown already in \cite{HV} that this is not
possible for the AVV triangle graph and, indeed it is possible,
using techniques subsequently developed,
to show that the naive commutation relations would
lead to all loops, even convergent ones, containing an odd numbers of
$\gamma_5$ 's vanishing \cite{BM}. This has created the awkward situation
where often the naive relations are used except when they are
known to give inconsistent results.

	A much more serious problem is connected with the violation
of chiral symmetry. Again in \cite{HV} it was noted that the
prescription Eq.\ref{eq:com} gave the correct result for the axial
anomaly, something that could not occur with a chirally-symmetric
regulator. Since this is a true anomaly, this is all to the good.
However, in addition to this ``good'' anomaly this regulator also
produces a number of ``bad'', spurious anomalies. These are spurious
in the sense that they can be removed by appropriate
renormalization. The appropriate renormalization requires
renormalization constants which themselves are not chirally
symmetric. This was shown long ago in \cite{TLT} and is
systematically discussed in Collin's book, \cite{JC}. There is no
fundamental problem with the procedure, but it is often very
laborious. What makes the situation especially treacherous is when
this is encountered in conjunction with dimensional {\it
renormalization}. This will in general not provide the appropriate
renormalization required. The additional finite renormalizations must
be put in explicitly ``by hand'', and they are sometimes overlooked.
This danger also arises in calculations which are finite for some
algebraic reason. An example of this will be given later.

In this regard, it is important to note the limitations of the oft cited work
of
Breitenlohner and Maison \cite{BM}. That paper carries through very
systematically the task of showing that dimensional renormalization is
a consistent renormalization procedure. They emphsize, however, that the
procedure does not preserve equations of motion and Ward identities when \gf\
is
present. See the discussion they give of their Theorem 2 which is limited to
the
case where there is no explicit dependence in the relations on the dimension
$d$. Such $d$-dependence, as pointed out by them, is the source of anomalies.
See  Bonneau \cite{GB} for a thorough discussion of this point. (Bonneau's
review also cites a large number of works on this topic. We do not attempt a
historical review here, but cite only those papers which directly bear on the
present work.)

In most cases these spurious anomalies can be avoided by using the naive
commutation relations. This was realized in some of the earliest applications
of the method, \cite{WB,BGL,CFH,GD}, and led to a series of attempts to
formulate rules for manipulating \gf\ in arbitrary dimension that essentially
produce the naive rules except in those instances where they are known to be
inconsistent and for which appropriate modifications are provided. For one
reason
or another none of these prescriptions have stuck. We do not propose to review
them here, but there are a number of reasons that may explain this. Some sets
of rules are not complete or, at least, it is not clear how they are to be
applied in general. In other cases, the rules require manipulation of
admittedly non-existent objects called \gf\, or they require radical
redefinition of the operation of taking the trace of a product of matrices
around a closed loop.

An instance of this latter approch is to be found in \cite{KKS}. This is
cited here because it was developed in response to a very interesting example
of the dangers of problems occuring in finite calculations mentioned above, and
because it illustrates the continuing difficulties in handling \gf\ \cite{BDG}
{}.
In that case, the flavor changing decay $Z \rightarrow b + \bar{s}$ is
calculated
at one loop order. Because of the basic structure of the Lagrangian this must
come out to be a finite, well-defined number. The authors of \cite{BDG}
compared the results of using Eq.\ref{eq:com} with the results of naively
commuting \gf\ and they found that they gave finite but different results. In
order to resolve this unsatisfaactory state of affairs they checked the Ward
Identities for the two cases and found that they failed for the
HV definition but were satisfied for the naive assumption. They conclude from
this that Eq.\ref{eq:com} ``does not appear to be a practicable \gf\
-prescription in amplitudes with open fermion legs as it produces a number of
spurious anomalies already at the one-loop level.'' In fact, application of the
procedure presented in Collins' book, \cite{JC} generalized to include several
generations of quarks, leads to the same result as does anticommuting \gf\ .
The
key ingredient to be added is operator mixing under renormalization: flavor
diagonal neutral currents can --- and do --- mix with flavor off-diagonal
neutral
currents. This is worked through explicitly in the Section 2.2. We draw
attention to this here to dispel any doubts that this may have caused regarding
the fundamental soundess of Eq.\ref{eq:com}. On the other hand, it is
undeniable
that the naive rules lead much more easily to the correct answer.

The objective of this paper is to provide a simple way of obtaining the correct
results when \gf\ rears its ugly head by a modification of the
regulator-renormalization procedure. These modifications will be {\em derived}
starting from Eq.\ref{eq:com} and, through a series of finite
renormalizations, result in rules that are simple to state, that do not involve
non-existent objects and do not require non-standard mathematics. The
regulating
procedure will be changed in such a way that no spurious anomalies arise and so
that renormalization by minimal subtraction gives consistent results. Happily
for most instances this leads to the naive rules in wide use. Necessarily,
closed loops with an odd number of \gf\ 's require a different procedure. The
one we have obtained we believe is new, simple to state and to use.

In Section 2 we will review and extend the results of \cite{TLT} for
non-singlet axial vector currents, including flavor changing currents and
operator mixing. This should make clear the issues and procedures. Section
3 will
deal with closed loops and the application to singlet axial vector currents.
Another source of spurious anomalies will be identified and dealt with, in
which
a closed loop is connected to another fermion line. A brief summary of
the procedures is given in Section 4. Some details are relegated to the
Appendix

\section{Open Lines}

\subsection{Non-singlet axial vector current}

The non-singlet axial current vertex is a good starting point
for the discussion of the method. It avoids the complexities of
closed loops with an odd number of \gf\ 's while at the same time
providing a physically important illustration of the issues.

The problems involved here all hinge on the following points: (1) in
calculating an open fermion line containing a \gf\ in one loop or
higher order one invariably meets the combinations
$\gamma_\alpha\gamma_5\gamma^\alpha$ or
$\gamma_\alpha\gamma_\lambda\gamma_5\gamma^\alpha$ where a sum over
$\alpha$ is implied. If the naive commutation relations were valid,
the \gf\ could be pulled out to the left or right and the sum would
be determined by vector vertex sum; this can be simply carried to an
arbitrary number of contractions such as one meets in multi-loop graphs.
However Eq.\ref{eq:com} gives a different result:
\begin{eqnarray}
\gamma_\alpha\gamma_5\gamma^\alpha = -4\gamma_5 + (d-4)\gamma_5
&\label{eq:con1}\\ \gamma_\alpha\gamma_\lambda\gamma_5\gamma^\alpha =
2\gamma_\lambda\gamma_5 - (d-4)\gamma_\lambda\gamma_5
&\label{eq:con2}
\end{eqnarray}
There are two things to note about these relations: first, the only
difference from the naive case is the sign of the terms carrying
the $(d-4)$ factor, where $d$ is the dimension of space-time;
second, the correction term is twice as big, relatively, for the
second equation as it is for the first. Both of these differences
will play havoc with Ward identities, which are algebraically valid
for the naive commutation relations. These differences will be
compounded by the multiple contractions that occur in higher order;
indeed, two contractions are required already in one loop, Fig.1.
(2) when these relations are encountered in a divergent graph or
sub-graph the poles that are produced at $d=4$ by the momentum integrations
combine with the $(d-4)$ factors above to produce a constant
difference between the two commutation relations proportional to
\gf\ and $\gamma_\lambda\gamma_5$, respectively.

At the one loop level, these differences are constants and so can
be viewed as a simple finite renormalization. If one used some kind
of mass-shell or momentum subtraction, these would be absorbed into
the infinite renormalization constants and never seen again. If, on
the other hand one simply subtracts off the pole at $d=4$, possibly
along with some fixed constants, the difference between the two
would remain in the nominal renormalized result. (An extra trap
occurs in the case of the Landau gauge which gives a finite result and
does not force a renormalization.) This of course cannot be and
the finite renormalization must be done if the axial vector current
is to obey the Ward identities required of it.

This result is very suggestive that the full axial vector vertex function, to
all orders in QCD calculated according to Eq.\ref{eq:com}, is a finite
renormalization of that calculated naively. Some time ago we reported
in \cite{TLT} results directed at
clearing up a problem of the generic type we are discussing here,
\cite{NW}. An inductive proof was given that, to all orders in QCD
for massless quarks
\begin{equation}
\Gamma_\mu^5 = z(\alpha_s)\Gamma_\mu \gamma_5
\label{eq:old}
\end{equation}
Here $\Gamma_\mu$ and $\Gamma_\mu^5$ denote, respectively, the vector
and axial vector current vertices calculated using Eq.\ref{eq:com}
with the standard $d$-dimensional $\gamma$-algebra and with the
divergences removed by minimal subtraction. $z(\alpha_s)$ is a finite
renormalization constant. Evidently the chiral
symmetry which is broken by the rules of Eq.\ref{eq:com} can be restored
by defining a renormalized axial current which is $1/z(\alpha_s)$ times the
minimally subtracted current. This result can obviously also be
obtained by calculating from the beginning with anticommuting \gf\ .

Eq.\ref{eq:old} is a very explicit way to express the result for the
massless theory. We will sketch the extension of this
inductive procedure to $m\neq 0$ in order to emphasize the important
issues. For any non-singlet vertex graph or subgraph $G$ with the \gf\
attached to an open line and {\em whose external legs are in $d=4$ }, denote
by $\tilde{G}$ the same graph evaluated using the naive commutation
relations; viz. one simply anticommutes \gf\ out of the graph naively
before doing the gamma algebra. \gf\ can be anticommuted out in either
direction, the result is the same.

Let $\Gamma_\mu^{5{(n)}}$ denote the axial vector vertex to
$n^{th}$-order in the QCD coupling $\alpha_s$ and $\widetilde{\Gamma}_\mu^{5
(n)}$ the corresponding sum of graphs using the anti-commuting
\gf\@. Because the term in the lowest order graph, Fig.1, proportional to
$m$ is convergent and because of Eq.\ref{eq:old} to order $n=2$ it is
obvious that
\begin{equation}
\Gamma_\mu^{5 (2)} = z_2(\alpha_s)\widetilde{\Gamma}_\mu^{5 (2)}
\label{start}
\end{equation}
This is used as the first step in the inductive argument based on
the Dyson-Schwinger equation illustrated in Fig.2. We do not wish
to belabor this standard kind of argument here. We just indicate that in
each of the blobs all subgraphs are assumed to be made finite by
minimal subtraction. The blob marked $K$ is two-particle irreducible.
(One can follow the argument of Bjorken and Drell, \cite{BD}. The
graphs involving two or three photon intermediate states, which they
dispose of by Furry's Theorem or by gauge invariance, do not enter
here because this is a non-singlet current.) Thus the $k$ integration
is not involved in any divergent subgraphs and  the only remaining
divergence is the overall integration. So if we differentiate once
 the sum of graphs represented by Fig.2, see \cite{CK}, the result
will be power-counting convergent and can be evaluated in $d=4$; in
particular $k$, the external momentum of the vertex can be taken to be
in $d=4$. We then use the induction hypothesis
\begin{equation}
 \Gamma_\mu^{5(m)} =
z_m(\alpha_s) \widetilde{\Gamma}_\mu^{5 (m)} \; \mbox{for $m<n$} \label{induc}
\end{equation}
to replace $\Gamma_\mu^{5(m)}$ by its tilded partner, anticommute
\gf\ through the $S_F$ and $K$ which are now in $d=4$ and conclude
that
\begin{eqnarray}
\Gamma_\mu^{5(n)} -
z_{n-1}(\alpha_s) \widetilde{\Gamma}_\mu^{5 (n)}= a_n
\alpha_s^{n}\gamma_{\mu} \gamma_5 & \mbox{$a_n$ a constant, or} \\
\Gamma_\mu^{5(n)} = z_n(\alpha_s) \widetilde{\Gamma}_\mu^{5 (n)}
\label{eq:new}
\end{eqnarray}
This is the desired extension of our earlier result for $m\neq
0$. It is important to emphasize that these results are valid only in
the limit that $d \rightarrow 4$ and for the external legs in $d=4$.
Otherwise additional ``evanescent'' operators come in to the relations.
See \cite{JC}. This caused the derivation to be somewhat fussier than
might at first sight seem necessary. It will cause further
difficulities in the discussion of closed loops.

This procedure allows us to perturbatively calculate $z(\alpha_s)$
should we wish to. However, it really isn't necessary. Rather we note
that exactly the same procedure applied to the vertex of the
pseudoscalar density yields
\begin{equation}
\Gamma^{5 (n)} = z_n^{5}(\alpha_s) \widetilde{\Gamma}^{5 (n)}
\label{eq:new'}
\end{equation}
Note $z_n \neq z_n^{5}$.
But then standard algebra shows that
\begin{equation}
(p'-p)^{\mu} \widetilde{\Gamma}_\mu^5(p',p)= \gamma_5 S^{-1}(p) +
S^{-1}(p') \gamma_5 - 2im\widetilde{\Gamma}^5(p',p) \label{eq:axial Ward}
\end{equation}
That is, $\widetilde{\Gamma}_\mu^5(p',p)$ is the properly normalized
axial vector vertex, and the finite renormalization that is required to
remove the spurious anomlies induced by \ref{eq:com} is automatically
achieved by using the naive commutation relations. This is, of course,
the conventional procedure, but we have justified it here is a way that
will be useful later.

One can argue quite generally that this must be the result: the two
different regulation schemes can only lead to finite renormalization.
In the case at hand the only dimension $3$ axial vector operator is the
the original current so it must be multiplicatively renormalized.
Because the minimal renormalization of
$\widetilde{\Gamma}_\mu^5(p',p)$ is the same as that of the conserved
vector current, it is renormalization invariant. Likewise with the
pseudoscalar density. Furthermore we see that the minimal
renormalization of $\widetilde{\Gamma}^5(p',p)$ is the same as the
vertex of $\bar{\psi} \psi$ and since $m\bar{\psi} \psi$ is
renormalization invariant in a mass independent renormalization scheme
like minimal subtraction \cite{SW}, so is $m$ times the renormalized
pseudoscalar
density. That is, the finite renormalizations of the two operators
$j_{\mu}^{5}$
and $j^5$  must be precisely those that are obtained by the naive commutation
relations in order that Eq.\ref{eq:axial Ward} be renormalization scale
invariant.

\subsection{Higher Order Weak Interactions}

The next most complicated situation arises when \gf\ occurs as an
internal vertex on an open line, as it does in higher order weak
interaction calculations. We do not propose to discuss the
renormalizability of $SU(2)_L\otimes U(1)$ here; rather we will examine
some important examples in low orders in the weak interactions. A very early
study of various \gf\ schemes in weak decays may be found in Marciano
\cite{WJM}.

The simplest and most obvious example of the need to carry out the
finite renormalization that is achieved by naive \gf\ commutation
rules is the vertex correction to single $W^\pm$ exchange, to which
must be added the unphysical charged scalar $\phi$ exchange, Fig.3. Because
of the different factors in Eq.\ref{eq:con1}and Eq.\ref{eq:con2} if they
are used without the finite renormalization the amplitude will contain an
unphysical term
\begin{equation}
Cg^2\frac{(m_u+m_d)(m_s+m_c)}{k^2-\xi M_{W}^{2}}
\end{equation}
where $C$ is a non-zero constant. The finite renormalization
constants for the two graphs are different and are such as to
compensate for this term.  Using naive \gf\ from the beginning avoids the
necessity of doing this. This is the simplest example which illustrates
that removal of infinities is not sufficient to guarantee the
consistency of the theory, and that attention must be paid to the
spectrum. \cite{DG}.

Henceforth in all graphs in this section we will assume that, for
QCD corrected vertex graphs and subgraphs, the required finite
renormalization has been carried out; i.e. within divergent
vertex subgraphs, naive \gf\ is used along with minimal
subtraction. If the graphs are overall divergent there will in general
be different renormalizations for the two schemes, and this will be our
focus here.

Let us start by examining the second order weak process mentioned in the
introduction, \cite{BDG}. Here the electroweak Ward identities will
force the choice of \gf\ commutation relations. Consider the set of
graphs Fig.4. All of these except for
the unphysical $\phi$  graphs are finite by power counting or by
the GIM mechanism \cite{GIM}, and so give finite results independent
of \gf\ commutation relation. The $\phi$  contribution to
wavefunction renormalization is, to this order, also independent of
\gf\ . All the problem then arises from $\phi$  exchange, the third graph,
because
it has a \gf\ buried inside a divergent graph. The difference between
the two procedures is easy to calculate since it comes only from the
divergent part. The result is
\begin{equation}
\Gamma_\mu
-\widetilde{\Gamma}_\mu=\frac{i}{128\pi^{2}}\frac{g^2}{cos\theta_W}
C_{bs}\gamma_\mu (1- \gamma_5)/2 \label{gamdif}
\end{equation}
where $C_{bs} = \sum V_{si}^{\dagger} V_{ib} m_i^2/M_W^2$; the
sum on $i$ goes over the quarks $u,c$ and $t$. At first sight it is not
obvious how to account for this difference because there is no zeroth
order flavor changing $Z$ coupling to apply a finite renormalization to.
The answer is, of course, that under Eq.\ref{eq:com} the axial vector
current $j_{\mu}^5$ is not (partially) conserved. Therefore it is
renormalized and mixes with other operators of the same dimension.
Indeed, the minimal subtraction is proportional to operators precisely
of the form $\bar{b} _L \gamma_\mu s_L $. What is needed here is an
additonal finite renormalization proportional to the same operator to
restore the chiral symmetry, just as in the first subsection.
Thus

\begin{equation}
j_\mu^{5}\rightarrow
j_{\mu
R}^{5}=j_\mu^{5}-Z_{ij}^{d}\bar{d_i}\gamma_\mu \frac{1-\gamma_5}{2}d_j-
Z_{ij}^{u}\bar{u_i}\gamma_\mu \frac{1-\gamma_5}{2}u_j
\end{equation}
By imposing $\partial^{\mu}j_{\mu R}^5 = \Sigma 2m_i^d \bar{d}_i
\gamma_5 d_i +  \Sigma 2m_i^u \bar{u}_i \gamma_5 u_i$ one easily
finds that %
\begin{equation}
Z_{bs}^{d}=\frac{i}{128\pi^{2}}\frac{g^2}{cos\theta_W} C_{bs}
\end{equation}
exactly the factor found by Barroso et al \cite{BDG} to be needed.
When this renormalization in applied
$\Gamma_\mu \rightarrow \widetilde{\Gamma}_\mu$, and so when this
formalism is carried through consistently it produces the correct
result. There is no denying, however, that care is required and that it
is much simpler {\em and correct} to apply the naive commutation
relations from the beginning.

We turn now to the process $b\rightarrow s+\gamma$, a problem in which
the \gf\ question has received much attention, \cite{bsgamma}. We
focus attention on the Green's function $\langle T(j_{\mu}(x) b(y) \bar{s}
(z)\rangle$ to lowest order in the weak coupling $g$ but, eventually, to
all orders in $\alpha_s$. The simplest graphs shown in Fig.5 are of
order $g^2 (\alpha_s)^0$. The wiggly lines refer to either the $W$
boson or to its unphysical charged scalar partner. (Note that the
correct hermitian axial vector vertex is $[ \gamma_{\alpha},
\gamma_5 ] /2$.) These graphs give the same result for the vertex
renormalization in either scheme because GIM \cite{GIM} makes the $W$
contribution
finite and \gf\ appears only external to the divergent loop in the $\phi$
contribution in (a) and (b). The $W$ contribution to (c) and (d) leads to an
off-diagonal mass renormalization which cancels between the two when the
external legs
are on shell. It is well-known that such terms can be removed by a linear
redefinition of the fields and have no physical consequences \cite{FKW}. For an
explicit realization of the redefinition appropriate to this case see
\cite{SB}; in
general, the left- and right-handed components are transformed differently:
\begin{equation}
\psi \rightarrow \sqrt{Z_0}\psi \; \; \mbox{and} \; \; \bar{\psi} \rightarrow
\bar{\psi} \sqrt{\bar{Z_0}}
\end{equation}
where $\bar{Z} = \gamma_0 Z^{\dag} \gamma_0$ and
\begin{equation}
Z_0 = Z_0^L \frac{1-\gamma_5}{2} + Z_0^R \frac{1+\gamma_5}{2}.
\end{equation}
$Z_0^{L,R}$ are matrices in flavor space; to the order we are working in $g$
the
diagonal elements can be taken to be unity. The transformations are not
necessarily
real; for that reason the redefinition disposes of off-diagonal mass
renormalization as well as wave function renormalization. To this order the
renormalization is pure imaginary (corresponding to a unitary
transformation which
rotates the off diagonal mass terms away).

This case differs from the previous case because the vector current
in conserved in either scheme; it does not depend on the \gf\
definition. Therefore we expect that there is no current
renormalization and the only change in the Green's function must
result from the quark propagator renormalization. We write the
Green's function as a matrix in $b$ and $s$ space
\begin{equation}
(S \Gamma_{\mu} S)_{bs} = S_{bb} \Gamma_{\mu bs} S_{ss} +
S_{bb} \Sigma_{bs} S_{ss} \Gamma_{\mu ss} S_{ss} +
S_{bb} \Gamma_{\mu bb} S_{bb} \Sigma_{bs} S_{ss}  .
\end{equation}
When projected on shell this gives $\Gamma_{\mu bs} + \Sigma_{bs}
S_{ss} \Gamma_{\mu ss} + \Gamma_{\mu bb} S_{bb} \Sigma_{bs}$ for the
matrix element for $b \rightarrow s + \gamma $.

Now go to order $\alpha_s$; Fig.6 shows the graphs relevant to our
problem. To these must be added graphs related by symmetry, those with $W
\rightarrow \phi$, graphs where $W$ and $g$ are disjoint as well as self
energy and
vertex counter terms. In graphs Fig.6b,g,h and i the \gf\ is anticommuted
naively
out of the subgraphs to compensate for the finite axial vector matrix
renormalization. The mass renormalization obtained in lowest order enters
in Fig.6c
and its partner (not shown). It is easy to show by partial fractions that this
yields $\sqrt{Z_0} _{bs} \Gamma_{\mu ss}^{(2)} + \Gamma_{\mu
bb}^{(2)}\sqrt{\bar{Z_0}}
_{bs}$. There are no other divergent subgraphs so the only remaining
difference can
come from the overall divergence of the sum of all the graphs. This has  the
form $\alpha_s g^2 (a_1 \gamma_\mu + a_1^5 \gamma_\mu \gamma_5)$. Gauge
invariance
ensures that the self energy difference is given by
\begin{equation} \alpha_s g^2
(-a_1 \gamma \cdot p -a_1^5 \gamma \cdot p \gamma_5 + \Delta m_1 + \Delta m_1^5
\gamma_5)
\end{equation}
Changes in the diagonal Green's functions do not enter the process $b
\rightarrow s +
\gamma$ at order $g^2$. This form is again equivalent to a linear field
redefinition at order $g^2 \alpha_{s}$. (The explicit form can be obtained from
the formulas in \cite{SB}, but we won't need them.)These differences will
cancel in the
physical amplitudes with $b$ and $s$ quarks on shell and so the two schemes
will give
the same result; there is no finite renormalization required.

In order to proceed to higher order it is necessary to determine
the change in the $gb\bar{s}$ vertex $\Gamma_{\mu}^g$ in going from one
scheme to
the other. One can see explicitly that to this order the
change is exactly the same as for the $\gamma b\bar{s}$ vertex.
(The set of graphs is similar to Fig.6 with the omission
of 6a and 6b and the addition of graphs containing the triple-gluon
vertex.) Using
induction, as usual, we assume that to any order $m<n$ that
\begin{eqnarray}
S^{(m)} & = & \sqrt{Z_m} \widetilde{S}^{(m)} \sqrt{\bar{Z_m}} \\
\Gamma_{\mu}^{(m)} & = & \sqrt{\bar{1/Z_m}}
\widetilde{\Gamma}_{\mu}^{(m)} \sqrt{1/Z_m}\\
\Gamma_{\mu}^{(m)g} & = & \sqrt{\bar{1/Z_m}} \widetilde{\Gamma}_{\mu}^{(m)g}
\sqrt{1/Z_m}
\end{eqnarray}
continuing with matrix notation. Then by the usual differentiation we
have that
\begin{equation}
\Gamma_{\mu}^{(n)} = \sqrt{\bar{1/Z_{n-1}}}
\widetilde{\Gamma}_{\mu}^{(n)}\sqrt{1/Z_{n-1}} + g^2
\alpha_s^n (a_n \gamma_{\mu} + a_n^5 \gamma_{\mu} \gamma_5) ,
\end{equation}
\begin{equation}
\Gamma_{\mu}^{(n) g} = \sqrt{\bar{1/Z_{n-1}}}
\widetilde{\Gamma}_{\mu}^{(n)g} \sqrt{1/Z_{n-1}} + g^2
\alpha_s^n (a_n^g \gamma_{\mu} + a_n^{g5}\gamma_{\mu} \gamma_5) ,
\end{equation}
\begin{equation}
\Sigma^{(n)}(p) = \sqrt{\bar{1/Z_{n-1}}}
\widetilde{\Sigma}^{(n)}(p) \sqrt{1/Z_{n-1}} + \sum_{m=0}^n g^2
\alpha_s^m(b_m \gamma \cdot p + b_m^5 \gamma \cdot p \gamma_5 +
\Delta m_m + \Delta m_m^5 \gamma_5). \label{eq:sigma}
\end{equation}
The terms in the sum for $m<n$ are fixed by the induction hypothesis:
\begin{equation}
\sum_{m=0}^{n-1} g^2
\alpha_s^m(b_m \gamma \cdot p + b_m^5 \gamma \cdot p \gamma_5 +
\Delta m_m + \Delta m_m^5 \gamma_5) = (\gamma \cdot p - m_s)
\sqrt{Z_{n-1}}_{bs} +
\sqrt{\bar{Z}_{n-1}}_{bs} (\gamma \cdot p - m_b);
\end{equation}
but then we can read off from Eq.\ref{eq:sigma} the terms of order
$\alpha_s^n$ in
these renormalization constants. The electromagnetic Ward identity implies that
$a_n = - b_n$ and $a_n^5 = - b_n^5$ and so to order $n$ the physical matrix
element
is unchanged.

We need to also argue that the $gb\bar{s}$ vertex change is the
same. Since the change is a constant mattix of the same form as the
photon vertex we can fix it by projecting on-shell. Then, because
the color current is conserved (or equivalently using BRS
identities)
\begin{equation}
(p'-p)^{\mu} \Gamma_{\mu bs}^g + \Sigma_{bs} S_{ss} \Gamma_{\mu
ss}^g  (p'-p)^{\mu} + (p'-p)^{\mu} \Gamma_{\mu bb}^g S_{bb}
\Sigma_{bs} = 0 .
\end{equation}
for $\gamma \! \cdot \! p' = m_{b}$ and $\gamma \! \cdot \! p  = m_{s}$. This
fixes the constants to be the same as in the photon vertex at order
$n$.

Thus the induction works and we find, as anticipated, that the only
change in the Green's functions results from wave function
renormalization. The consequence of this is that the same result
will be given by either scheme, to any order in QCD; no finite
renormalization enters. This result is not surprising. There was
considerable reason to
anticipate the result from the many explicit calculations to two
loop order for the effective four fermion theories, \cite{bsgamma}.
It is useful to work it through, however, to show that the {\em form} of
the difference between the two schemes is as anticipated.

The conclusion is that, since one must treat \gf\ naively within
the divergent axial vector vertex subgraphs and since it is so much simpler
otherwise---and the results are identical---one should certainly use naive
\gf\ for
all open lines.

The weak corrections to flavor {\em diagonal} processes to order
$g^2$ are  more complicated to discuss. First, one must take into
account the corrections to the external on shell wavefunctions and
show that these will yield the same physical result. This is
straightforward. One must also consider the closed fermion loops
that arise for the diagonal case. These will be discussed in the
next section.

\section{Closed Loops}

In this section, we will examine several classes of closed loop
graphs. In all of them the graphs will contain one and only one
closed fermion loop to which one or more axial vector or pseudoscalar
vertices are attached. The fermions in the loops will always be
massive. Arbitrary gluonic corrections to these loops are allowed
and, in particular, the \gf\ may appear inside a divergent
subgraph. All self-energy and vertex subgraphs are taken to be
renormalized by minimal subtraction. Starting, as we are, from
the \gf\ definition of 't Hooft and Veltman \cite{HV}, axial
vector and pseudoscalar vertices require additional finite
renormalization as discussed in Section 2. The equivalence of
this to using naive \gf\ commutation relations that was derived
there is valid only when the external fermion legs are in $d=4$,
and so the use of that procedure for divergent subgraphs within
closed loops must be justified or modified.

The classes we will examine are (1) loops which are
superficially convergent, (2) divergent loops corresponding to
renormalization parts : AA or PP bubbles and VAA triangle,
(3) weak corrections to closed loops,
(4) anomalous triangle and box diagrams, and (5) graphs constructed
by attaching the closed loop to an open fermion line. Class (4) is very
special:
consistency of gauge theories require that the triangles within the gauge
theory be
non-anomalous so the anomaly can occur consistently only for non-gauged,
external
currents such as interpolating fields for hadronic matrix elements. An
important
example is the operator corresponding to the longitudinal polarization in deep
inelastic lepton-nucleon scattering \cite{polarization}. Although it is
special, it
has become a touchstone of any method of treating \gf\ and so
cannot be ignored, even though it demands much more careful
attention than the other cases. Class (5)  represents another potential
source of spurious anomalies. Completeness requires that the
method be adapted to this class as well.

The broadest class of convergent loops are those connected to
more than four external currents and are two and three gluon
irreducible. Graphs of this type are superficially
convergent and the fermion loop is powercounting convergent as
well. Thus, after doing the subgraph subtractions, the
corresponding momenta can be taken to be in $d=4$ \cite{JC} and
any required finite renormalization of axial vector or pseudoscalar
vertex subgraphs
can be achieved by treating \gf\ naively within the subgraph and, {\it a
fortiori}, \gf\ can be moved naively around the loop. The only
proviso is that $\gamma$-matrices within other divergent subgraphs
must be contracted within that subgraph and not across the \gf
\space. To fail to observe this proviso would run afoul of the
inconsistency mentioned in Section 2, and give the incorrect
renormalization of that subgraph.

Turning to class (2), let us look at the bubble graphs containing non-singlet
\gf\ vertices. There are four of these: each vertex can be either axial
vector or
pseudoscalar. We denote those calculated using the HV definition by
$\Pi_{\mu \nu}^{AA}$
and $\Pi^{PP}$, etc.. Correspondingly we use the notation  $\tilde{\Pi}_{\mu
\nu}^{AA}$ and $\tilde{\Pi}^{PP}$ for the same quantities calculated using the
naive commutation relations; there is no ambiguity in the latter.
By differentiating three times with respect to the fermion mass or
the external momentum one creates an overall power counting convergent
integral so that the results of  Section 2 can be used to
renormalize the divergent \gf\ subgraphs and we have
\begin{eqnarray}
\Pi_{\mu \nu}^{AA}(q) & = & z^2 \widetilde{\Pi}_{\mu \nu}^{AA}(q) +
T(g_{\mu \nu} q^2 - q_{\mu} q_{\nu}) + L q_{\mu} q_{\nu} + M m^2
g_{\mu \nu} \\
\Pi_{\mu}^{PA}(q) & = & zz_5\widetilde{\Pi}_{\mu}^{PA}(q) + Rq_{\mu}m^2 \\
\Pi^{PP}(q) & = & z_5^2\widetilde{\Pi}^{PP}(q)+ S q^2 + N m^2 .
\end{eqnarray}
In addition to the desired vertex renormalization this procedure
generates extra terms, as shown, with $T,L,M,R,S,N$ constants. These
terms correspond precisely to the ambiguity of defining the product
of local operators at the same point of the form $\partial_{\mu}
\partial_{\nu} \delta ^{(4)}(x-~y), \linebreak  m^2 g_{\mu \nu}\delta ^{(4)}
(x-~y)$ etc.\cite{Bogie}. It is very similar to the ambiguity
discovered many years ago by Chanowitz, Furman and Hinchliffe
\cite{CFH}. The resolution of this ambiguity is part of the definition of
the axial
current. If the bubbles occur inside a graph, as corrections to the weak boson
propagator, all of these constants except $T$ must be chosen so that
$\Pi_{\mu \nu}^{AA} \rightarrow \widetilde{\Pi}_{\mu \nu}^{AA}, \Pi_{\mu}^{PA}
\rightarrow  \widetilde{\Pi}_{\mu}^{PA}$ and  $\Pi^{PP} \rightarrow
\widetilde{\Pi}^{PP}$ in order to avoid
unphysical poles in graphs such as in Fig.7. This is the same point made at the
beginning of Sec. 2.2. $T$ is fixed by requiring that, for $m=0$ the vertex and
axial vector bubbles are the same to preserve chiral symmetry \cite{TLT}.

Next consider the VAA graph of Fig.8a. This
is known to be free of true anomalies. However, it is not superficially
convergent
and so if it is calculated, alternatively, with Eq.\ref{eq:com} or with
naive \gf\
different results are obtained, the difference being of the form
\begin{equation} A (p-p')_{\lambda} g_{\mu \nu} + B \{ (p+p')_{\mu}
g_{\lambda \nu} - (p+p')_{\nu} g_{\mu \lambda}\}
\label{eq:VAA}
\end{equation}
where $A$ and $B$ are finite constants. As in the last case,
this difference corresponds to the amibiguity in the
definition of three local currents at a singlepoint of the form \linebreak
$g_{\mu \nu}\delta^{(4)}(x-~y) \partial_{\lambda}\delta^{(4)}(y-~z)$ etc.
To resolve this ambiguity, consider the interesting case where this
arises when each boson is in
$SU(2)_L$. See Fig.8b. Then by Furry-like arguments only the VVV and
three VAA triangles survive, proportional to $\epsilon_{abc}$.
Naively all four are equal but by Eq. \ref{eq:VAA} the three
VAA are different from the VVV and and the sum of the differences is equal to
\begin{equation} (A+B)\{ (p-p')_{\lambda}g_{\mu
\nu}+(k-p)_{\nu}g_{\lambda \mu} +(p'-k)_{\mu}g_{\nu \lambda}\} .
\end{equation}
This is exactly the form of the bare triple-boson vertex and so the
difference between the two is just a finite vertex
renormalization, analogous to that seen in Section 2. Evidently,
again the naive \gf\ leads to the chiral symmetric equality of VVV
and VAA as $m \rightarrow 0$. As in the last example, this renormalization
must be
carried out if these triangle occur internally in order to avoid unphysical
poles.
This evidently remains true when arbitrary numbers of gluons are attached to
the
fermion loop.  This justifies the folk theorem, that even numbers of \gf\ can
be
removed from a closed loop by anticommutation; indeed, to avoid the need
for finite
renormalizations to restore the spuriously broken chiral symmetry, vital to the
consistency of the $SU(2)_L$ gauge theory, they {\it must} be anticommuted
naively and
removed from the loop.

Let us next consider the weak corrections to closed loops such as shown in
Fig.9,
again with an arbitrary number of gluon corrections to the loop. The $W$
vertices will
yield loops containing 0, 1, or 2 \gf\  's. One can see that the loops with
one \gf\
are convergent: as before, for the photon graph the divergence must be
proportional to
$K \epsilon_{\lambda \mu \nu \alpha} (p-p')^{\alpha}$, but electromagnetic
current
conservation requires that $K=0$. Symmetry requires that the gluon graph
have the form
$f_{abc} \epsilon_{\lambda \mu \nu \alpha} (p_1 + p_2 + p_3)^{\alpha}$, but
this is
identically zero. Similar
arguments, using the same physics, can be presented for those graphs with box
subgraphs coupling $W^+ W^-$ to $g g$ in the first instance and four gluons
in the
second.

For loops containing two \gf\ 's Furry's theorem gives zero for the photon
graph. The
three gluon graph is non-zero but has a divergence proportional to $f_{abc}
V_{\lambda
\mu \nu}$, the usual triple boson coupling. This leads simply to a finite
renormalization of the strong coupling constant. Thus, when the weak
interactions are
taken into account the strong coupling will differ by an amount proportional to
$g^2 \alpha_s$ in the two schemes.  This is a familiar situation and is not
aproblem,
Of course, one must be sure to use the same scheme in {\it all} graphs in
order to
preserve BRS symmetry, and to note that the numerical value of $\alpha_s$
will depend
on the scheme as well.

The AVV case where the axial current is a singlet under the gauge
group of the vector currents is historically the most interesting.
For many currents of physical interest, the fermion content of the
theory is such that when the loops with all the different fermions
are summed the result is power-counting convergent
\cite{DG,Georgi}. Thus, for these cases the procedures given for convergent
loops in
the first paragraph of this section apply and there is no true anomaly. We
will not
dwell on this but turn at once to the case of an anomalous singlet current.
It is
imperative that the \gf\ procedure deal naturally with this case in order
for it to be
considered satisfactory. The AVVV case falls into the same category. The
long history
of this subject begins, of course, with Adler and Bell and Jackiw
\cite{ABJ}. More
recent papers that are especially pertinent to this subject are
\cite{Leveille},
\cite{Akhoury} and \cite{Bos} where the graphs of Fig.10 are calculated.

As in all cases we begin with the fermion loops calculated using the
original 't Hooft-Veltman definition of \gf\ , the loop integration and the
$\gamma$-algebra being done in $d$-dimensions. The spurious anomaly studied in
Section 2 arises here in two loop order for the first time and must be
removed by
the finite renormalization $z(\alpha_s)$, see Fig.10a and 10l. Because the
triangle is
not power-counting convergent it is not possible to take the loop momenta
$k$ in $d=4$
and to take the limit of the (renormalized) axial vector vertex to $d=4$.
Therefore one
cannot use Eq.\ref{eq:new} to automate the required finite renormalization
as we did for
the open lines. Notice, by the way, that triangles where the axial vector
vertex is
replaced by a pseudoscalar vertex are power counting convergent and
Eq.\ref{eq:new'} can
be used to automate the finite renormalization $z_5(\alpha_s)$. Our goal is
to modify
the regulation so that the spurious anomaly in the axial vector triangle is
automatically renormalized away. We will proceed on the basis of two
properties of
these graphs. The first is that
\begin{equation}
\lim_{m \rightarrow \infty} \Gamma_{\lambda \mu \nu}^5(p,p',m) = 0
\end{equation}
This follows from BRS invariance. See the Appendix for  an inductive
demonstration of
this property. The second property is that all the power counting divergent
subgraphs
which are not compensated by the usual self-energy and vertex minimal
subtractions
contain the complete fermion loop. Thus, the combination $\Gamma_{\lambda \mu
\nu}^5(p,p',m) - \Gamma_{\lambda \mu \nu}^5(p,p',M)$ is power counting
convergent. We
may then evaluate this combination making use of this property. There are many
possible ways to do this. The only essential one is that we may use
Eq.\ref{eq:new}  to
achieve the required finite renormalization of divergent axial vector
subgraphs of the
fermion loop; the additional changes induced by this replacement are
independent of
$m$ and will cancel in the above combination. Beyond that, for example,
one could evaluate the graphs using naive \gf\ ; there would result
ambiguities in each term in the sum depending on how the various $\gamma$
's that
appear in the loop are contracted with respect to the \gf\ but these
ambiguities will
cancel in the sum: because the derivative of these graphs with respect to $m$
is
convergent the ambiguities are independent of $m$. We will return to this
flexibility later.Thus we have
\begin{equation}
\Gamma_{\lambda \mu \nu}^5(p,p',m) = \lim_{M \rightarrow
\infty}(\Gamma_{\lambda \mu
\nu}^5(p,p',m) - \Gamma_{\lambda \mu \nu}^5(p,p',M)) = z \lim_{M \rightarrow
\infty}(\widetilde{\Gamma}_{\lambda \mu \nu}^5(p,p',m) -
\widetilde{\Gamma}_{\lambda
\mu \nu}^5(p,p',M)) \label{eq:final}
\end{equation}
It is important to bear in mind that $ \lim_{M \rightarrow \infty}
\widetilde{\Gamma}_{\lambda \mu \nu}^5(p,p',M)$ exists only for the gauge
invariant
sums---that is, there will be logs of $M$---and not graph-by-graph. It will
generate
the set of graphs arising from  the gauge non-invariant vertices
$Z_A(\alpha_s) \epsilon_{\lambda \mu \nu \alpha} A^{\mu a}
\partial_{\alpha} A^{\nu a}$
and  $Z_D(\alpha_s) \epsilon_{\lambda \mu \nu \rho} f_{abc} A^{\mu a}
A^{\nu b} A^{\rho
c}$, which will automatically be the set of counter terms required to
restore gauge
invariance to the sum in Eq.\ref{eq:final}. Because the loops that contain a
pseudoscalar vertex are powercounting convergent, we obtain
\begin{eqnarray} z^{-1}iP^{\lambda} \Gamma_{\lambda \mu \nu}^5(p,p',m) & =
 & \lim_{M \rightarrow \infty} P^{\lambda} (\widetilde{\Gamma}_{\lambda \mu
\nu}^5(p,p',m) -  \widetilde{\Gamma}_{\lambda \mu \nu}^5(p,p',M)) \\
& = & 2m \widetilde{\Gamma}_{\mu \nu}^5(p,p',m) + A(\alpha_s) \Gamma_{\mu
\nu}^{F
\tilde{F}}(p,p') \\
& = & 2mz_5^{-1}\Gamma_{\mu \nu}^5(p,p',m) + A(\alpha_s) \Gamma_{\mu \nu}^{F
\tilde{F}}(p,p') , \label{eq:chiral Ward}
\end{eqnarray}
where $P=p+p'$ and
\begin{equation}
A(\alpha_s) \Gamma_{\mu \nu}^{F\tilde{F}}(p,p') = - \lim_{M \rightarrow \infty}
2M \widetilde{\Gamma}_{\mu \nu}^5(p,p',M)
\end{equation}
is the true anomaly.
$\Gamma_{\mu \nu}^{F\tilde{F}}$ denotes the $F^a\widetilde{F}^a$
fermion--anti-fermion vertex function \cite{Adler}.
This method constructively determines the coefficient $A(g^2)$. Note that
we do not use
or address the Adler-Bardeen theorem here \cite{AB}.

The precise definition of $\widetilde{\Gamma}$ has been left free so far.
Any ambiguity
is removed in the difference taken in Eq.\ref{eq:final}. One can choose to do
more-or-less as one pleases; the only essential is that the \gf\ within the
divergent
axial subgraph be treated naively as in Eq.\ref{eq:new}. It is good to make
use of this
freedom to split the Green's functions into two pieces: one which obeys the
naive
chiral Ward identity, without any anomaly, and one piece which is ``pure''
anomaly;
i.e. it is the corresponding Green's function of the counter terms
$Z_A(\alpha_s) \epsilon_{\lambda \mu \nu \alpha} A^{\mu a}
\partial_{\alpha} A^{\nu a}$
and  $Z_D(\alpha_s) \epsilon_{\lambda \mu \nu \rho}
f_{abc} A^{\mu a} A^{\nu b} A^{\rho c}$, as we will now see.

 When a closed loop, even a convergent one, is attached to an open
fermion line the potential for a new source of spurious anomalies is
encountered which
cannot be dealt with by simply by modifying \gf\ commutation rules. These are
anomalies for which the analog of Eq.\ref{eq:chiral Ward} are violated so
that the
underlying renormalized operators do not satisfy the anomalous chiral Ward
identity
\begin{equation} \partial^{\mu} j_{\mu}^5 = 2m j^5 + A(\alpha_s) F^a
\widetilde{F}^a  \label{eq:acWi}
\end{equation}
without further finite renormalization of $j_{\lambda}^5$, the analog of the
non-singlet renormalization in Section 2.1. They occur because, even for
convergent
loops, the analogs of the relation Eq.\ref{eq:chiral Ward} are valid only
for $p,p'$ in
4-dimensions and in the limit $d \rightarrow 4$; corrections proportional
to $d-4$ can
give a finite result when combined with the overall divergence of the graph.
The
simplest example is shown in Fig.11 and it occurs already for Abelian
theories: to the
minimally subtracted graph (a) must be added the finite renormalization of
order
$\alpha_s^2$ (b)  appropiately chosen in order that the divergence yield
graph (c). This
was recognized and explicitly calculated by Larin \cite{Larin} and used to
calculate
$\alpha_s^2 $ corrections to the Ellis-Jaffe sum rule, \cite{polarization}.  He
calculated the required finite renormalization constant by insisting that the
renormalized operators satisfy this equation.

Another approach is to modify the regulation
procedure in a way analogous to Section 2, so that Eq.\ref{eq:acWi} is
automatically satisfied with all operators renormalized by minimal
subtractions.
There are probably a variety of ways of doing this which could, in principle,
in principle generate finite renormalizations to both $j_{\mu}^5$ and
$F\tilde{F}$. We
propose the following way, which seems simple to state and to use. To any
order, the
complete gauge invariant set of graphs $\Gamma$ breaks up into sets,
labeled by index
\{$i$\}; each graph in the set have exactly the same gluon configuration
and the set is
defined by all possible insertions of the axial vector vertex onto the
loop. (For
example, Fig.10 the sets are (a,b,c,d), (e,f), (g,h,i),(j) and (k).) Were
it not for the
divergences, these sets would individually satisfy the naive chiral Ward
identity. For
each set, choose a fixed vector vertex---say the marked index in
Fig.10---and naively
anti-commute \gf\ from the axial vector vertex to that fixed vertex. By
using the
convergent combination Eq.\ref{eq:final}, one does not change
$\Gamma_{\lambda \mu
\nu}^5(p,p',m)$ except for multiplying by a finite factor analogous to $z$
for the
vector vertex chosen. We won't need to calculate that factor, though one
could, and
that would provide an alternative to the procedure we will describe. Call
the function
so defined $\Gamma^N$, because it manifestly satifies the naive chiral Ward
identity
{\em even for $d \neq 4$ and for the external momenta with components
outside the first
four dimensions.} This is very important because now this decomposition can
be used when
this Green's function appears as a subgraph of a divergent graph. The limit $M
\rightarrow \infty$ is to be taken {\em before} this is attached to an open
fermion
line, because it is being used to generate the two- and three-gluon
counter- terms. The
limit does not exist for graphs with an open fermion line \cite{Adler,CWZ}.

This is the decomposition that we sought and one could just use it with the
HV \gf\,
calculating the new finite vector renormalization constant, but we would
like to bring
it more into the spirit of what we have been doing. Separate the primary
loop from the
rest of the graph by cutting all gluon lines which connect it to the
external vector
lines. Call this function $L^N_{\lambda \cdots}(P,q,\cdots,m)$ where $P$ is the
momentum exiting through the axial vector vertex, $q, \cdots$ stands for
the momenta
of the various gluons which connect the loop to the rest of the graph and
$m$ is the
fermion mass in the loop. Evidently for each set $i$
\begin{equation}
iP^{\lambda}L^N_{\lambda \cdots}(P,q,\cdots ,m)=2mL^N_{\cdots}(P,q,\cdots, m)
\end{equation}
implies that
\begin{equation}
L^N_{\lambda \cdots}(0,q, \cdots, 0) = 0
\end{equation}
where the various other momenta $q, \cdots$ are arbitrary non-vanishing
vectors in
$d$-dimensions. In particular for the two-gluon function $p+p'=P=0$ but
$p'-p \neq 0$.
By forming  %
\begin{equation}
L^N_{\lambda \cdots}(P,q,\cdots,m) = L^N_{\lambda
\cdots}(P,q,\cdots,m) -  L^N_{\lambda \cdots}(0,q, \cdots, 0)
\label{eq:zero mass sub}
\end{equation}
we have created a manifestly power-counting convergent form for $L^N$ and
all the graphs into which it falls will also be power-counting convergent.
We may
therefore evaluate these sets of graphs in $d=4$, including the fermion loop.
Therefore naive \gf\ commutations can---and should---be used to avoid axial
vector and pseudoscalar subgraph renormalizations, always bearing in mind
the proviso
mentioned earlier regarding the divergent self-energy and vector vertex
subgraphs
of convergent loops.

This gives us the first piece of our decomposition; the other piece, the purely
anomalous piece is calculated by taking the  limit $M \rightarrow \infty$ of
the
piece just calculated. As emphasized earlier, for the complete, gauge
invariant set
the limit exists and is finite. (These steps can be followed through
explicitly in
the simplest non-trivial example shown in Fig.10 by using the Tables in
Akhoury and
Titard \cite{Akhoury}.)

In all of these cases the primary closed loop can be evaluated in $d=4$,
after doing
the divergent vertex and self-energy insertions, and so \gf\ can be treated
naively
throughout, especially within the divergent axial vector vertex subgraph. By
bringing \gf\ back to where it started, in both the axial vector and the
pseudoscalar
triangle, the finite vector renormalization we encountered in the course of the
derivation is undone. Indeed, all the steps involved there were just to
show that this
works; they can now be forgotten and we can calculate the Green's functions
in the
form we started from, provided we make the additive corrections
Eq.\ref{eq:zero mass
sub} and Eq.\ref{eq:final}. This ``solves'' the \gf\ problem for closed
loops. To
achieve this it has been very important to contrive that \gf\ appears only in
convergent closed fermion loop integrals. Other tricks might be used, but the
pitfalls are many.

If  the function is attached to an open line, because the graph so formed is
log
divergent, replacing the original graph by this decomposition is equivalent
to a finite
renormalization of $\Gamma^5_{\lambda}$. This is exactly the finite
renormalization
(e.g. $a'_2$ of Fig.11, the finite renormalization calculated by Larin
\cite{Larin})
required to maintain Eq.\ref{eq:acWi}: the first piece is completely
convergent and so
gives the first term of Eq.\ref{eq:acWi}, and $P$ contracted into the
second piece is
explictly equal to the anomaly and so minimal subtraction of each maintains
that
relation.

Because of the form of this decomposition, there should be no difficulty in
extending it to graphs with additional closed fermion loops, bearing in
mind that $M$
as used above is not a global regulator, but is a tool for determining the
non-gauge
invariant renormalization constants required. For example, consider the
graphs in
Fig.12 which are a sample of the simplest gauge invariant set of two
fermion loop
graphs. It is simple to show that $\lim_{M \rightarrow \infty}$ of the sum
of all
these graphs vanishes. There are two finite renormalizations required when
using
Eq.\ref{eq:com}: Fig.12c requires a finite renormalization because of the
divergent
axial vector vertex and Fig.12a requires one for the reasons we have just
seen. They
can both be avoided by using our technique of separating the subgraphs into two
pieces. Note that the appropriate counter-terms for the fermion loops in
Fig.12a and
12b are determined by taking the  $\lim_{M \rightarrow \infty}$ of those
loops before
the final pair of integrations are done. Because the graphs are not overall
power
counting convergent making these replacements for the subgraphs will lead
to a result
that can differ from our starting point by a constant.
Just as before this constant can be determined by the $\lim_{M \rightarrow
\infty}$ of
the {\em full} graphs calculated in this manner.

This method for graphs involving anomalous closed loops is reasonably
simple to state
and it seems reasonably direct to implement, but only experience will show
if it is
more efficient than the completely equivalent method of using
Eq.\ref{eq:com} and
calculating the various finite renormalization constants which are needed to
compensate for the spurious anomalies introduced thereby. Our principal
goal here has
been to address  a broad range of \gf\ problems and reduce
them to a coherent approach within the minimal subtraction approach to
renormalization which avoids the algebraic complications of the non-covariant
Eq.\ref{eq:com} and the necessity (and dangers) of additional finite
renormalizations.

The conclusion is that, with the proper care with regard to cases (4) and
(5), the HV
scheme with the finite renormalizations required by the symmetries of the
theory
yields results which are {\it equal} to those obtained by naive commutation
relations.

\section{Summary}
Starting on the basis of the consistency of the definition of \gf\
given in the original discussion of dimensional regularization by
't Hooft and Veltman \cite{HV}
\begin{equation}{
\gamma_5 = i\gamma_0\gamma_1\gamma_2\gamma_3\;}
\label{def}
\end{equation}
and the necessity of making a series of finite renormalizations to
remove the spurious anomalies in the chiral symmetry equations that
arise from this definition, we have developed a procedure for
evaluating Feynman graphs that contain \gf\ and using
dimensional regularization with minimal subtraction that is simple
and avoids the need for the finite renormalizations. The procedure
is both simple to state and simple to implement:

(1) For open fermion lines containing axial vector or pseudoscalar
vertices, one should calculate the subgraphs containing the
\gf\ as if \gf\ anticommutes with {\it all} $\gamma$
matrices, i.e. ``naive'' \gf\ . The \gf\ may then be freely
anticommuted along the fermion line. This will ensure, without
further renormalization, that non-singlet currents satisfy
\begin{equation}
(p'-p)^{\mu} \Gamma _{\mu} ^5 (p', p) = \gamma _5 S_F^{-1}(p) +
S_F^{-1}(p') \gamma _5 - 2im \Gamma ^5 (p',p)
\end{equation}
where all quantities are renormalized by minimal subtraction. This
is, of course, the commonly used procedure for this type of graph.
We emphasize that here it is {\it derived} from the consistent
starting point stated at the start of this section and is not offered
as a fundamentally alternative procedure.

(2) If \gf \space  occurs within a convergent loop in a convergent
graph it should be treated as if it satisfies the naive
anticommutation relations. This ensures that the renormalizations
that are required for  divergent axial vector or
pseudoscalar vertex subgraphs that may occur within the convergent loop
are properly accounted for. The only caveat is that all the $\gamma$
{}~-algebra contractions within other divergent subgraphs---self-energy
and vector vertices---must be done internally to that subgraph and
not contract across the \gf\ . The same is true for divergent loops---
generalized bubble, triangle or box graphs---which contain an even
number of \gf\ 's; this takes account of the overall renormalization
required in addition to restore the chiral symmetry between vector and
axial vector Green's functions and to ensure that the unphysical gauge
dependent poles in vector boson propagators properly cancel against
those of the corresponding unphysical scalars.

(3) Superficially divergent loops containing a single axial vector
vertex require special attention. We consider only color singlet
axial vector currents so the potentially problematic cases are the
triangle and box graphs with all their radiative corrections. For
definiteness, we consider the other currents to be colored vector
currents. Other cases are simpler and easily deduced from this
result. If when the various fermions which can circle in the loops
are added together the resulting sum is convergent, as it is for
the electroweak axial vector current in the standard model of
fermion doublets, then as in (2), \gf\ should be treated naively.

The interesting case is when the sum is not power-counting
convergent and a true anomaly occurs. There are many ways to proceed;
one can define a unique way by insisting on writing the gauge invariant
set as a sum of a term which obeys the naive chiral Ward identity
and a term which is pure anomaly. This can be achieved in the
following way: for any graph with a single fermion loop attached to
the axial vector current and any number of gluons, subtract from it
the same graph with everything the same except that in the fermion
loop the total momentum entering the axial vector  vertex $P=0$ and the
mass of the fermion is set to zero. For that combination, every
subgraph, including the whole graph, containing the fermion loop is
superficially convergent. Therefore it can be calculated using naive \gf\ .
Furthermore, it satisfies the naive chiral Ward identity.  The complete, gauge
invariant result is obtained by summing all terms of a given order of the above
combination and subtracting from it the finite function obtained by
taking the limit $M \rightarrow \infty$. This generates the
complete set of graphs with the vertices
$Z_A(\alpha_s) \epsilon_{\lambda \mu \nu \alpha} A^{\mu a}
\partial_{\alpha} A^{\nu a}$
and  $Z_D(\alpha_s) \epsilon_{\lambda \mu \nu \rho} f_{abc} A^{\mu a}
A^{\nu b} A^{\rho c}$ required as counter-terms to restore gauge
invariance.

(4) When a closed loop, convergent or divergent, containing a
\gf \space is attached to an open fermion line creating a graph which is
overall divergent there is a new source for potential spurious anomalies
in the sense that the anomalous chiral Ward identity arising in case (3)
\begin{equation} \partial^{\mu} j_{\mu}^5 = 2m j^5 + A(\alpha_s) F^a
\widetilde{F}^a  \label{eq:finale}
\end{equation}
is not preserved by minimal subtraction and
an additional finite renormalization of $j_{\mu}^5 $ is required. Use of
the above decomposition avoids this problem because the convergent
combination, since it satisfies the naive chiral Ward identity, has
good asymptotic behaviour in the integration momenta and so for this
term the potentially divergent graphs are in fact power-counting
convergent and contraction with $P$ gives just the first term
in Eq.\ref{eq:finale}. The remaining term is precisely the
anomaly so that the minimal subtraction of those terms is identical to
minimal subtraction of the anomaly-two fermion vertex.

  These procedures have been derived to
all orders in QCD and to second order in the electroweak axial coupling
for graphs with an open fermion line, a single closed fermion loop and a
closed loop connected to an open fermion line.   There should be no
problem in extending this to graphs with additional fermion loops. We
are not aware of any problem in going to higher order in the electroweak
coupling, but we have not examined it in detail.

\bigskip
{\bf Acknowlegements:} Thanks are due to W. Marciano, R. Akhoury, J.
Collins, R. Kauffman, J. Korner, A.Soni for helpful comments and special
thanks are due to Eduardo deRafael who called my attention to this
problem many years ago and to Scott Willenbrock whose persistent
questions reawakened my interest much more recently.
\newpage

\appendix
\section{Appendix}
The following argument is needed in Section 3 and is included here
for completeness; it may be well known but we haven't found a
source for this specific argument although the general approach
follows, as does much of this work Collins' book \cite{JC}. We show that
a gauge invariant set of graphs for AVV or AVVV containing only the
fermion loop to which the axial vector current is attached must
vanish as the fermion mass $m$ goes to infinity.

Start with the simple loop, properly symmetrized in the
external vector legs. As $m\rightarrow \infty$ this amplitude must go
to a dimensionless function of $m/\mu$ times $\epsilon_{\lambda \mu
\nu \alpha} (p-p')^{\alpha}$ or  $\epsilon_{\lambda \alpha \beta
\gamma}$ for the AVV or AVVV case respectively. The coefficients
 do not depend on the external momenta because two derivatives of
the first case or one of the second with respect to the external
momenta must vanish as $m \rightarrow \infty$. BRS symmetry requires
that the divergence of the amplitude with respect to one vector
index must vanish when it is contracted into the polarization vector
of the other {\em on-shell} vector legs. These tensors do not have
that property and so, because the coefficient of the tensor does not
depend on the momenta, it must vanish identically. Evidently, a
simple loop with more than three gluon lines attached will vanish by
power counting.

For an inductive argument, assume that to
some order $n$ it has been shown that a gauge invariant sum of graphs
with two or three gluon legs and one axial
vector current vanish as $m \rightarrow \infty$. To order $n+1$
a subset of the graphs have the external gluons attached directly
to the fermion loop. By the above argument these must go to
$f(m/\mu)\epsilon_{\lambda \mu\nu \alpha}(p-p')^{\alpha}$ or
$g(m/\mu)\epsilon_{\lambda \alpha \beta \gamma}$ for the AVV or AVVV
case respectively. For short, in the following discussion, we will
call this ``going to a constant''. For the remaining graphs, examine
any subgraph which contains the fermion loop and has $r$ external
legs. If $r>3$ these subgraphs vanish as $m \rightarrow \infty$.
For $r=2,3$ add together all the graphs which are obtained by
replacing the subgraph in question by another subgraph of the same
order and with the same number of external legs. The sum of these
subgraphs is gauge invariant and by the induction hypothesis it
vanishes as $m \rightarrow \infty$. Subgraphs which have ghost
lines connecting the fermion loop to the external gluons are
power counting convergent and so vanish as $m \rightarrow
\infty$. Therefore only the case where all internal lines of the
graph are of order $m$ as $m \rightarrow \infty$ are possibly
non-vanishing and the sum of this subset of graphs may go to a
constant. However, when all the graphs of order $n+1$ are added
together the sum of these constants must vanish. Therefore, the
induction goes through and the gauge invariant set must vanish as
$m \rightarrow \infty$ at any order  .

\newpage
\thispagestyle{empty}
\section*{Figure Captions}
\newcounter{fig}
\begin{list}%
{Fig.\arabic{fig}}{\usecounter{fig} \setlength{\rightmargin}{\leftmargin}}

\item The lowest order QCD correction to the axial
vector vertex.
\item  The Dyson-Schwinger equation for the non-singlet
axial vector vertex.
\item  Illustration of the requirement for finite
renormalization of the $W$ and $\phi$ couplings.
\item  The lowest order diagrams for $Z \rightarrow b\bar{s}$.
\item The lowest order diagrams needed for $b \rightarrow s \gamma$. The
wiggly lines
represent $W$ or $\phi$ in every graph.
\item  A sample of the relevant two loop
contributions to  $b \rightarrow s \gamma$. See text.
\item  Illustration of the need for finite
renormalization of the bubble graphs coupling $W$'s and
$\phi$'s.
\item  (a)The general VAA triangle and (b) the
non-vanishing triangles coupling three $SU(2)_L$ currents.
\item  An order $g^2$ contribution to $\gamma$gg and ggg coupling.
\item  The order $\alpha_s^2$ contributions to the axial
vector coupling to two gluons. The vertex marked with a heavy dot is a possible
selected vector vertex. To these graphs must be added those with the two
external
gluons interchanged to maintain the Bose symmetry.
\item  The simplest graphs illustrating the need for a finite
renormalization $a'_2$
of graphs connecting a closed loop to an open fermion line.
\item  Sample of three loop anomaly graphs containing two fermion loops.
The constant
$a''_2$ consists of two pieces: one from the vertex renormalization
in (c) and one from the renormalization coming from Fig.11 in (a).
\end{list}

\bigskip

\end{document}